\documentclass[11pt]{article}
\usepackage{moriond,epsfig}

\def\be{\begin{equation}}
\def\ee{\end{equation}}
\def\bea{\begin{eqnarray}}
\def\eea{\end{eqnarray}}

\begin{document}
\vspace*{4cm}
\title{TOP QUARK MASS MEASUREMENTS AT THE TEVATRON}

\author{ MARTIJN MULDERS \\ (on behalf of the CDF and D\O\ collaborations)}

\address{Fermi National Accelerator Laboratory, Batavia, IL 60510, USA
  \\ \vspace{.3 cm}}

\maketitle\abstracts{In the year 2004 several milestones in the 
measurement of the top quark mass were reached. The D\O\ 
collaboration published a significant
improvement of their Run I measurement of the top quark mass, and both
Tevatron experiments released preliminary measurements based on Run II 
data sets collected in the period 2002-2004. The preliminary Run II results
presented here do not yet surpass the current world average in
precision, but this is expected to change soon. With larger
data sets ready to be analyzed, a better understanding of
the Run II detectors and improved analysis methods, 2005 promises to
be a remarkable year for Top physics.}

\section{Introduction}

The recent publication of the improved Run I measurement of the top
mass by D\O~\cite{ME} was exciting for two reasons. First of all it demonstrated how
much improvement in measurement precision could be achieved using a more
advanced analysis technique like the Matrix Element method. Secondly, it was a reminder of how
little we yet know about the properties of the top quark and that new
experimental information about the top quark can have big implications 
for electroweak fits in the Standard Model. The current (Run I only)
world average value for the top quark mass is $178.0 \pm 4.3 $
GeV$/c^2$. In the coming years the measurements of CDF and D\O\ combined
should lead to a precision of about 2 GeV. Together with expected
improvements in the measurement of the W boson mass this will allow
to further constrain the Higgs boson mass to a relative precision of
approximately 30\%, as discussed elsewhere in these proceedings~\cite{hayes}.

Since the start of Run II both CDF and D\O\ have recorded more than
600 pb$^{-1}$ of data, already 5 times the Run I luminosity. The
preliminary results presented here are based on fraction of the
recorded data ranging from 160 to 230 pb$^{-1}$.

\section{Run II Top mass results}

In $p\bar{p}$ collisions with $\sqrt{s} = 1.96$ TeV at the Tevatron,
top quarks can be produced via the strong interaction in $t\bar{t}$
pairs, or as single top quarks through the weak interaction. Single top
production is predicted to have a lower cross-section and a more
challenging event signature, and has not yet been observed at the time
of this conference. For the Top mass measurement therefore only top
pair events are used. Each top quark decays immediately to a $W$ boson
and a $b$ quark, and the $W$ bosons decay either hadronically or
leptonically, giving rise to 3 possible decay channels: di-lepton,
lepton+jets and all-jets. 

An overview of recent $t\bar{t}$ cross-section results from the CDF
and D\O\ experiments in all three of the above final states is given
elsewhere in these proceedings~\cite{nielsen}. In both collaborations 
several top mass analyses are being developed in the di-lepton and lepton+jets 
decays channels, mostly based on very similar event selections. 
No preliminary Run II results in the all-jets channel have been 
presented so far. 

A complete and up-to-date overview of ongoing Run II analyses can be found on the
collaborations' public results web pages~\cite{cdfpub,d0pub}. A description of all 
analyses is outside the scope
of these proceedings. Below a few of the analyses
are briefly described in order to highlight some important 
aspects of the top mass measurement.

\subsection{Final states with two leptons plus jets}

The striking signature due to the presence of two leptons in the final
state allows for a relatively pure selection of top events, typically
with a signal-to-background ratio of 4/1. The main challenge however
is to fully reconstruct the kinematics of the final state, which are 
underconstrained due to
the presence of two neutrinos. Different approaches exist to add an
extra constraint to the system, and see for which value of the top
mass the observed events are most likely.

\begin{figure}
\psfig{figure=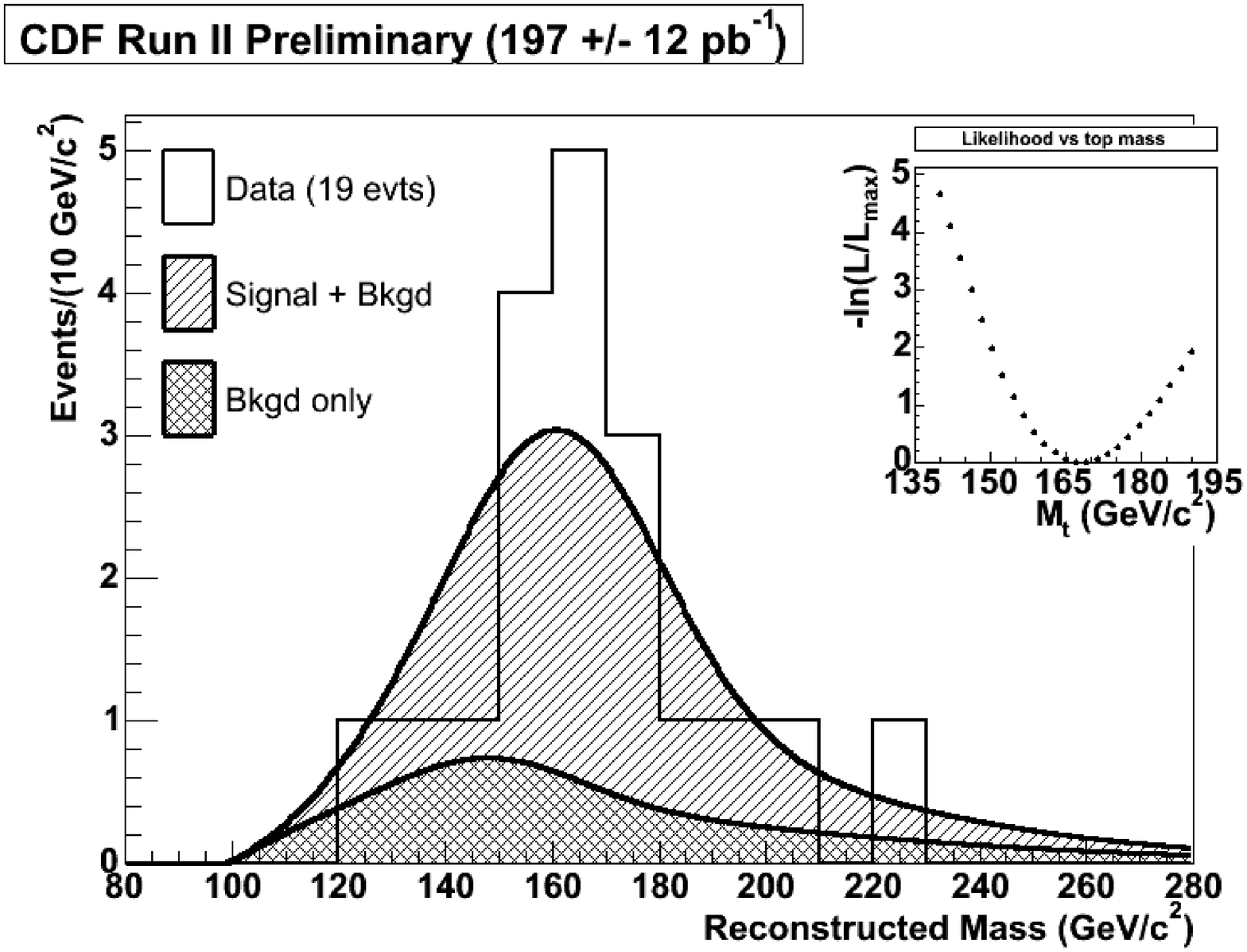,height=2.2in}\psfig{figure=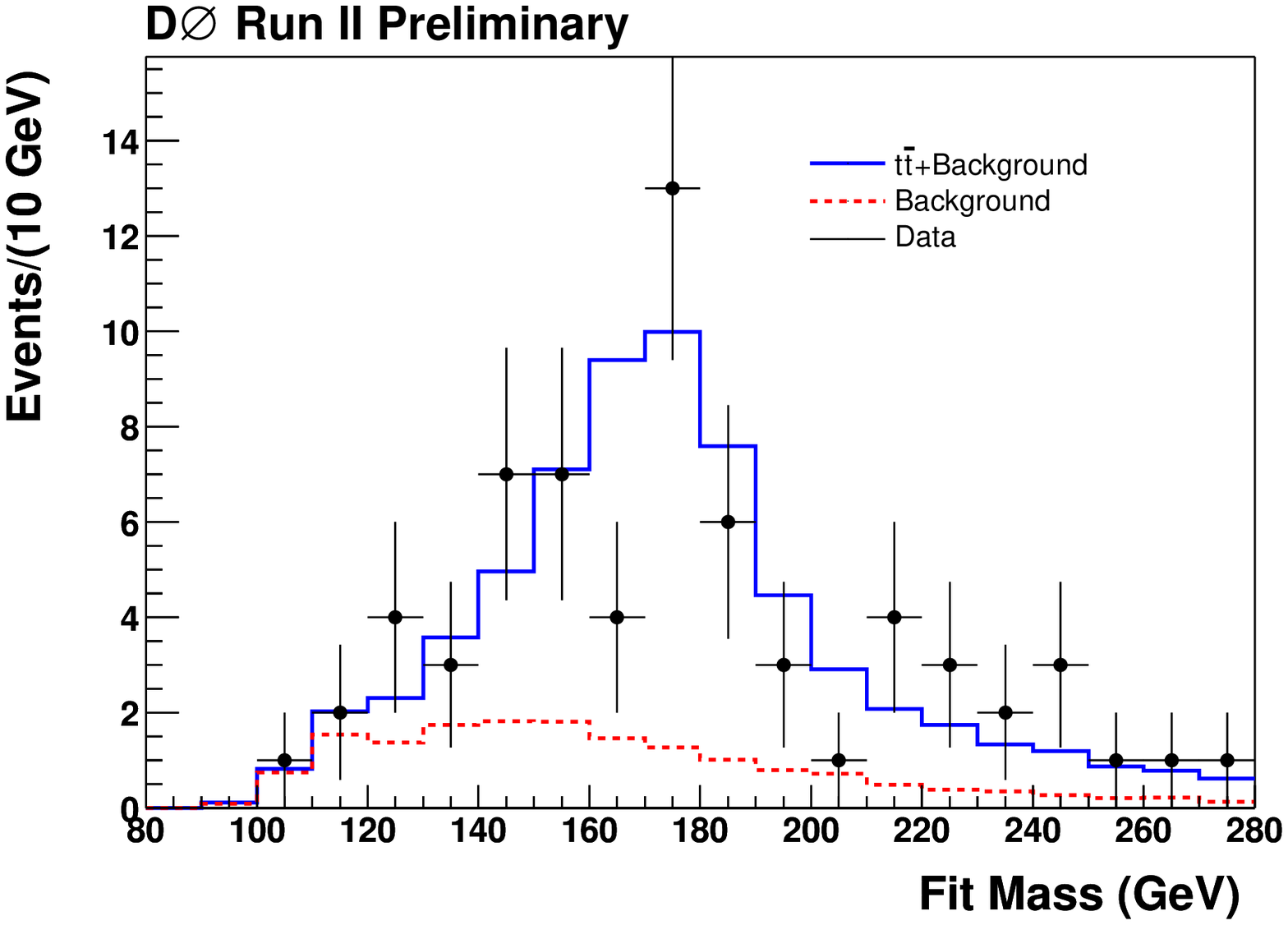,height=2.2in}
\caption{Reconstructed mass distributions for the CDF di-lepton
  neutrino weighting analysis (left), and the D\O\ Template method with
  b-tagging (right). 
\label{fig:plots}}
\end{figure}

In Table~\ref{tab:overview} several Run II analyses are listed with
their preliminary results. Currently the most precise result was
obtained by CDF with the neutrino weighting analysis using a loosened
lepton identification (one lepton + one isolated track), optimizing
the statistical precision by using a higher efficiency (and slightly
lower purity) selection. In this method the rapidities of both
neutrinos are used as extra constraints, and a weight as function of
the top mass is calculated by integrating over all possible rapidity
values and comparing the reconstructed missing transverse momentum with
the observed momentum imbalance using a Gaussian resolution. For
each event the top mass value which leads to the highest weight is
plotted and fitted using Monte Carlo Templates, as shown in
Figure~\ref{fig:plots}.  

\begin{table}[t]
\caption{Overview of preliminary Run II top mass results\label{tab:overview}}
\vspace{0.4cm}
\begin{center}
\begin{tabular}{|l|c|l|}
\hline
&  data set (pb$^{-1}$) & top mass (GeV$/c^2$) \\
\hline
di-lepton channel & & \\
\hline
 & &  \\ CDF neutrino-weighting                                      & 200 &
168.1 $^{+11}_{-9.8}$ (stat) $\pm$ 8.6 (sys) \\
CDF M$_{\rm reco}$ Template + $t\bar{t}$ $p_z$              & 194 &   176.5 $^{+17.2}_{-16.0}$ (stat) $\pm$ 6.9 (sys) \\
CDF M$_{\rm reco}$ Template + $\phi$ of $\nu_1$ and $\nu_2$ & 194 &
170.0 $\pm$ 16.6 (stat) $\pm$ 7.4 (sys) \\
D\O\ Dalitz and Goldstein                                     & 230 &
155 $^{+14}_{-13}$ (stat) $\pm$ 7 (sys) \\
 & &  \\ \hline
lepton+jets channel & & \\
\hline
 & &  \\ CDF Template with b-tagging & 162-193 &  177.2 $^{+4.9}_{-4.7}$ (stat) $\pm$ 6.6 (sys) \\
CDF Multi-Variate Template  & 162     &  179.6 $^{+6.4}_{-6.3}$ (stat) $\pm$ 6.8 (sys) \\
CDF Dynamic Likelihood      & 162     &  177.8 $^{+4.5}_{-5.0}$ (stat) $\pm$ 6.2 (sys) \\
D\O\ Ideogram                 & 160     & 177.5 $\pm$ 5.8 (stat) $\pm$ 7.1 (sys) \\
D\O\ Template topological     & 230     & 169.9 $\pm$ 5.8 (stat)
$^{+7.8}_{-7.1}$ (sys) \\
D\O\ Template with b-tagging  & 230     & 170.6 $\pm$ 4.2 (stat) $\pm$ 6.0 (sys) \\
 & &  \\ \hline
\end{tabular}
\end{center}
\end{table}

\subsection{Final states with one lepton plus jets}

While the lepton+jets channel benefits from a higher branching ratio, it
suffers from significant backgrounds from $W$+jets and non-$W$
multi-jet events. 

Since only one neutrino is present the final state can be fully
reconstructed. Some analyses use a constrained kinematic fit to
further improve the
measurement of lepton and jets beyond detector resolution. The CDF
Dynamic Likelihood Method (DLM) follows a different approach, similar to
the D\O\ Matrix Element method~\cite{ME}; transfer functions are derived from
Monte Carlo simulation describing the jet energy resolution. These 
functions are subsequently used in a multi-dimensional integration
over phase space calculating the likelihood that the event is compatible
with matrix elements describing top pair production and decay. 

In order to reconstruct the invariant mass of the top decay products,
a choice has to be made to assign jets and lepton to the corresponding
top or anti-top quark. In a lepton+jets event 12 ways exist to do
this assignment. Some analyses take only one jet assignment per event
in consideration. The CDF Dynamic Likelihood Method and the D\O\
Ideogram analysis include all possible jet assignments in the fit. 

The CDF and D\O\ template methods use an overall fit of Monte Carlo templates to
the data in order to extract the mass. The CDF Dynamic Likelihood
Method and D\O\ Ideogram analysis derive an event-by-event likelihood to
maximize the statistical information extracted from each event. The
Ideogram method also includes the hypothesis that the event could be
background, weighted according to an estimated event purity.

Both experiments apply b-tagging in some of the top mass
analyses. One advantage of
b-tagging is to strongly reduce the backgrounds. A second advantage of
b-tagging for the top mass measurement in the lepton+jets channel is
the reduction of the number of possible jet assignments in the case
that one or two jets are b-tagged. The CDF Template analysis combines
the 0-tag, 1-tag and double tagged event samples in the fit to
optimize the statistical precision. D\O's first top mass analysis with
b-tagging uses events with at least one tag, which applied to a data
set of 230 pb$^{-1}$ leads to the most precise preliminary Run II top 
mass result presented so far. Figure~\ref{fig:plots} shows the fitted
mass for the lowest-$\chi^2$ solution for the b-tagged D\O\ Template
analysis, compared to the Monte Carlo prediction.

An overview of the current preliminary results is shown in Table~\ref{tab:overview}.


\section{Prospects for the Top mass measurement}

In all results reported here the dominant component of the
systematic uncertainty is the uncertainties related to the jet
energy scale. In the last year a lot of work has been done to improve
the calibration of the reconstructed jet energies. CDF reports an
improvement of a factor two or more in jet energy scale uncertainties
compared to a year ago. Similar improvements are expected in D\O. This
will have a direct effect on the systematic uncertainties
quoted. 

Further improvements in understanding the Jet Energy Scale can come
from performing an in-situ calibration of the light-jet energy scale using
the jets from the hadronic decay of the $W$ in the same $t\bar{t}$
events used to measure the top mass, and from studies in progress aimed at
determining the b-jet energy scale from data. 

Other systematics that are being studied are the modeling of initial
state and final state gluon radiation in the $t\bar{t}$ Monte
Carlo.

Very soon both experiments hope to present preliminary results with
updated jet energy scale and an integrated luminosity of more than 300
pb$^{-1}$.  

All together the prospects are very good for having new top mass results
this year with a precision comparable to or better than the current
world average for each of the Tevatron experiments. This will open the
door to an exciting new area of top physics to be further explored in
the coming years at the Tevatron.

\end{document}